\documentclass[aps,prl,twocolumn,superscriptaddress,showpacs,showkeys]{revtex4-1}

\usepackage{subfigure}
\usepackage{graphicx}
\usepackage{dcolumn}
\usepackage{bm}

\usepackage{rotating}
\usepackage{amssymb}

\def\be{\begin{equation}} 
\def\ee{\end{equation}}
\def\bea{\begin{eqnarray}} 
\def\eea{\end{eqnarray}}
 
\newcommand{\cdhz}{\mbox{$^{110}$Cd} }
\newcommand{\pdhz}{\mbox{$^{110}$Pd} }
\newcommand{\cdhzp}{\mbox{$^{110}$Cd$^{+}$} }
\newcommand{\pdhzp}{\mbox{$^{110}$Pd$^{+}$} }

\newcommand{\obb}{0\mbox{$\nu\beta\beta$-decay} } 
\newcommand{\zbb}{2\mbox{$\nu\beta\beta$-decay} }
\newcommand{\ema}{\mbox{$\langle m_{\nu} \rangle $} }

\begin{document}


\title{$Q$-Value and Half-Lives for the Double-Beta-Decay Nuclide \pdhz}


\author{D. Fink}
\affiliation{Max-Planck-Institut f\"ur Kernphysik, 69117 Heidelberg, Germany}
\affiliation{Fakult\"at f\"ur Physik und Astronomie, Ruprecht-Karls-Universit\"at, 69120 Heidelberg, Germany}
\affiliation{CERN, 1211 Geneva 23, Switzerland}

\author{J. Barea}
\affiliation{Departamento de F\'isica, Universidad de Concepci\'on, Casilla 160-C, Concepci\'on, Chile}

\author{D. Beck}
\affiliation{GSI Helmholtzzentrum f\"ur Schwerionenforschung GmbH, 64291 Darmstadt, Germany}

\author{K. Blaum}
\affiliation{Max-Planck-Institut f\"ur Kernphysik, 69117 Heidelberg, Germany}
\affiliation{Fakult\"at f\"ur Physik und Astronomie, Ruprecht-Karls-Universit\"at, 69120 Heidelberg, Germany}

\author{Ch. B\"ohm}
\affiliation{Max-Planck-Institut f\"ur Kernphysik, 69117 Heidelberg, Germany}
\affiliation{Fakult\"at f\"ur Physik und Astronomie, Ruprecht-Karls-Universit\"at, 69120 Heidelberg, Germany}

\author{Ch. Borgmann}
\affiliation{Max-Planck-Institut f\"ur Kernphysik, 69117 Heidelberg, Germany}
\affiliation{Fakult\"at f\"ur Physik und Astronomie, Ruprecht-Karls-Universit\"at, 69120 Heidelberg, Germany}

\author{M. Breitenfeldt}
\affiliation{Instituut voor Kern- en Stralingsfysica, Katholieke Universiteit Leuven, 3001 Leuven, Belgium}

\author{F. Herfurth}
\affiliation{GSI Helmholtzzentrum f\"ur Schwerionenforschung GmbH, 64291 Darmstadt, Germany}

\author{A. Herlert}
\altaffiliation{Present address: FAIR GmbH, Planckstr.1, 64291 Darmstadt, Germany}
\affiliation{CERN, 1211 Geneva 23, Switzerland}

\author{J. Kotila}
\affiliation{Center for Theoretical Physics, Sloane Physics Laboratory,Yale University, New Haven, Connecticut 06520-8120, USA}

\author{M. Kowalska}
\altaffiliation{Present address: CERN, 1211 Geneva 23, Switzerland}
\affiliation{Max-Planck-Institut f\"ur Kernphysik, 69117 Heidelberg, Germany}

\author{S. Kreim}
\affiliation{Max-Planck-Institut f\"ur Kernphysik, 69117 Heidelberg, Germany}

\author{D. Lunney}
\affiliation{CSNSM-IN2P3/CNRS, Universit\'{e} de Paris Sud, 91406 Orsay, France}

\author{S. Naimi}
\altaffiliation{Present address: SLOWRI Team, Nishina Accelerator-based Research Center, RIKEN, 2-1 Hirosawa, Wako, Saitama 351-0198, Japan}
\affiliation{CSNSM-IN2P3/CNRS, Universit\'{e} de Paris Sud, 91406 Orsay, France}

\author{M. Rosenbusch}
\affiliation{Institut f\"ur Physik, Ernst-Moritz-Arndt-Universit\"at, 17487 Greifswald, Germany}

\author{S. Schwarz}
\affiliation{NSCL, Michigan State University, East Lansing, Michigan 48824-1321, USA}

\author{L. Schweikhard}
\affiliation{Institut f\"ur Physik, Ernst-Moritz-Arndt-Universit\"at, 17487 Greifswald, Germany}

\author{F. \v Simkovic}
\affiliation{Department of Theoretical Physics, Comenius University, 84848 Bratislava, Slovak Republic}

\author{J. Stanja}
\affiliation{Institut f\"ur Kern- und Teilchenphysik, Technische Universit\"at, 01069 Dresden, Germany}

\author{K. Zuber}
\affiliation{Institut f\"ur Kern- und Teilchenphysik, Technische Universit\"at, 01069 Dresden, Germany}



\begin{abstract}
The \pdhz double-beta decay $Q$-value was measured with the Penning-trap mass spectrometer ISOLTRAP to be $Q = 2017.85 \left(64\right)$\,keV. This value shifted by 14\,keV compared to the literature value and is 17 times more precise, resulting in new phase-space factors for the two-neutrino and neutrinoless decay modes. In addition a new set of the relevant matrix elements has been calculated. The expected half-life of the two-neutrino mode was reevaluated as $1.5\left(6\right)\times10^{20}$\,yr. With its high natural abundance, the new results reveal \pdhz to be an excellent candidate for double-beta decay studies.
\end{abstract}


\maketitle

\begin{figure} [t]
\includegraphics[width=0.49\textwidth]{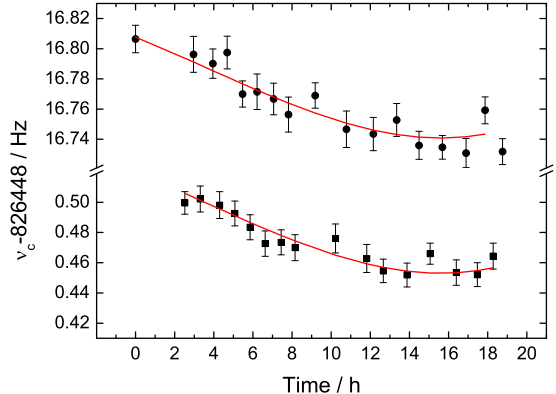}
\caption{Cyclotron frequencies of  \mbox{$^{110}$Cd$^{+}$} (circles, top) and of  \mbox{$^{110}$Pd$^{+}$} (squares, bottom) using the Ramsey excitation pattern 50-600-50\,ms. The solid lines are simultaneous fits to both isobars (see text).}
\label{fig:Poly_RamsS}
\end{figure}

The recent results on neutrino oscillations \cite{hosaka2006,adamson2008,abe2008,aharmim2010} have revolutionized our understanding of the role played by neutrinos in particle physics and cosmology, in particular by proving that neutrinos have a finite mass.  In the quest for a detailed understanding of the neutrino itself, the rare process of double-beta decay offers the most promising opportunity to probe the neutrino character and to constrain the neutrino mass \cite{Avignone2008,Rodejohann2011}. In contrast to neutrino oscillations, which violate the individual flavor-lepton number while conserving the total lepton number, the process of neutrinoless double-beta decay (0\mbox{$\nu\beta\beta$-decay}) violates total lepton number and is as such, forbidden by the Standard Model of particle physics.  Moreover, unlike the observed neutrino-accompanied double-beta decay (2\mbox{$\nu\beta\beta$-decay}) process, the \obb process would imply that the neutrino is a Majorana particle, i.e., its own antiparticle.\\
While the decay spectrum of the \zbb is continuous, the experimental signal of the \obb represents the sum energy of the two electrons at the decay $Q$-value. The expected half-life of the \obb is extremely long and hence, very small event rates are expected. In addition, a high accuracy (below 1\,keV) is desirable to properly identify the signal with respect to background. Furthermore, a well-known $Q$-value allows a precise determination of the phase space of the half-lives of the double-beta decay modes.\\
The decay rates of both decay modes are strong functions of the $Q$-value. \obb scales with the fifth power of the $Q$-value and \zbb with the eleventh power. Eleven nuclides ($^{48}$Ca, $^{76}$Ge, $^{82}$Se, $^{96}$Zr, $^{100}$Mo, $^{110}$Pd, $^{116}$Cd, $^{124}$Sn, $^{130}$Te, $^{136}$Xe, and $^{150}$Nd \cite{Zuber2004}) are considered as potential \obb candidates, having high enough $Q$-values (above 2\,MeV) for relatively short half-lives, which are accessible within the experimental limits and natural abundances for feasible experiments. Among these, \pdhz has the second highest abundance, making it a particularly interesting case. Furthermore, \pdhz is an excellent test candidate for the ``single-state dominance hypothesis'' (SSDH) \cite{Civitarese1998, Domin2005, Moreno2009}, which predicts that the double-beta decay rate is dominated by a virtual two-step transition through just one single intermediate $1^+$ state. Nevertheless, the experimental knowledge about the double-beta decay of \pdhz is still very limited. Recently, first experimental limits on excited state transitions were published \cite{Lehnert2011}. However, \pdhz has the highest $Q$-value uncertainty of all candidates mentioned above.\\
In this Letter, we report the first direct mass comparison between \pdhz and the double-beta decay daughter nuclide $^{110}$Cd. The new $Q$-value is shifted by almost 14\,keV to the literature value \cite{Audi2003} and has a 17 times smaller uncertainty. Furthermore, the absolute masses of \pdhz and \cdhz were measured with high precision within the same measurement campaign. We also report new calculations for the phase-space factors of the $\beta^-\beta^-$-double-beta decay modes based on the new $Q$-value. In addition, the relevant nuclear matrix elements have been calculted as well as the resulting half-lives of the neutrino-accompanied and neutrinoless double-beta decay.\\
The measurements were performed with the Penning-trap mass spectrometer ISOLTRAP \cite{Mukherjee2008} at the CERN-ISOLDE facility. ISOLTRAP determines the masses by measuring the cyclotron frequency $\nu_c = q B/(2 \pi m)$ of the corresponding ions with mass $m$ and charge $q$ in a magnetic field $B$. For the present 
\begin{figure} [t]
\includegraphics[width=0.284\textwidth]{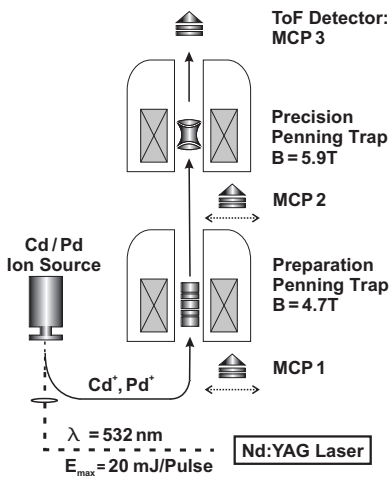}
\caption{Scheme of ISOLTRAP as used for the present measurements. The atoms are ionized with the ablation ion source, trapped in the first trap for purification and transferred to the second trap to measure their cyclotron frequency.}
\label{fig:ISOLTRAP}
\end{figure} 
off-line-experiment, a laser ablation ion source for the production of carbon-cluster reference-mass ions \cite{Blaum2002} was modified to deliver singly charged $^{110}$Pd and $^{110}$Cd ions. In short, a frequency-doubled Nd:YAG laser was focused with a typical pulse power density of $5\cdot10^7$\,W/cm$^2$ on either a carbon, a palladium or a cadmium foil for desorption and ionization. All three targets were mounted on a rotary sample holder for rapid element selection.\\
After the ablation from the sample, the ions were guided to the preparation Penning trap (see Fig.\ref{fig:ISOLTRAP}), where they were centered and purified by buffer-gas cooling \cite{Savard1991}. They were then transferred to the precision Penning trap where the time-of-flight ion-cyclotron resonance (TOF-ICR) detection technique \cite{Koenig1995} was used to determine their true cyclotron frequency $\nu_c$. Two different measurement sets using two different Ramsey excitation schemes \cite{George2007a,George2007b,Kretzschmar2007} were performed to increase the accuracy. The first excitation scheme (30-840-30\,ms) consisted of two rf-pulses of 30\,ms separated by a 840\,ms waiting period. For each resonance curve 
\begin{figure} [t]
\includegraphics[width=0.49\textwidth]{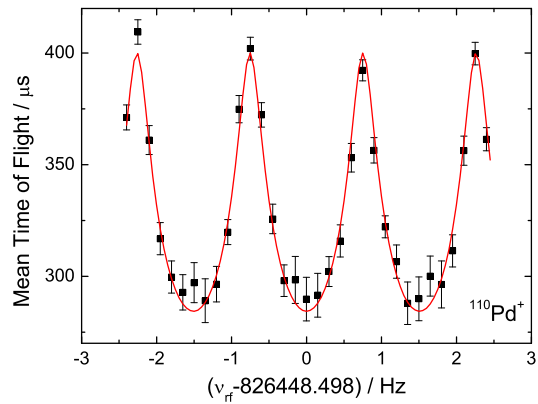}
\caption{Time-of-flight resonance as a function of the excitation frequency for \pdhzp for the Ramsey excitation scheme 50-600-50\,ms with a fit of the expected theoretical line shape (solid line) \cite{Kretzschmar2007}.}
\label{fig:RamseyPd}
\end{figure}
(Fig.\,\ref{fig:RamseyPd}), data were taken for about 20 minutes. In total, 17 such resonances of \cdhzp and 16 resonances of \pdhzp were recorded alternately. For the second set of measurements, a 50-600-50\,ms Ramsey excitation scheme was applied with the same duration of data-taking per resonance and the same number of resonances. 
The $Q$-value of the double-beta decay of the mother nuclide with mass $m_m$ to the daughter nuclide with mass $m_d$ is given by the mass difference $Q = m_m - m_d$, which in turn can be written as a function of the frequency ratio $r = \nu_d/\nu_m$ and the electron mass $m_e$:
\be \label{eq:Qvalue}
Q = m_m - m_d = \left( \frac{\nu_d}{\nu_m} - 1 \right) (m_d - m_e).
\ee
Figure\,\ref{fig:Poly_RamsS} shows the resonance frequencies for the 50-600-50\,ms Ramsey excitation scheme as a function of time of the measurement campaign with typical statistical uncertainties on the order of $\sigma(\nu_c)/\nu_c = 1 \cdot 10^{-8}$. In addition, different systematic uncertainties have to be taken into account: the presence of contaminating ions, the time-dependent magnetic-field changes, the mass-dependent systematic effect, and the residual systematic uncertainty of ISOLTRAP \cite{Kellerbauer2003}. A countrate-class analysis applied to the individual frequency measurements confirmed that there were no contaminating ions or space-charge effects present \cite{Kellerbauer2003}. The \cdhzp and \pdhzp data from Fig.\,\ref{fig:Poly_RamsS} follow the same trend, mainly given by a variation of the magnetic field strength in time. Following Bradley \textit{et al.} \cite{Bradley1999}, the frequency ratio $r$ and thus the mass ratio between the two ion species is deduced by fitting simultaneously a pair of polynomials
\begin{equation}
	 \label{eq:Polynom1}
	\nu_1(t)=c_0+c_1\cdot t+c_2\cdot t^2+...+c_n\cdot t^n\quad ,
\end{equation}
\begin{equation}
	 \label{eq:Polynom2}
	\nu_2(t)=r\cdot(c_0+c_1\cdot t+c_2\cdot t^2+...+c_n\cdot t^n)
\end{equation}
to the datasets of \pdhzp and \mbox{$^{110}$Cd$^{+}$}. Here, the frequency ratio $r$ entered as one of the fit parameters, while the other fit parameters represent the behavior of the magnetic field strength in time. Polynomials (Eq.\,(\ref{eq:Polynom1}) and Eq.\,(\ref{eq:Polynom2})) of different orders were tested with smallest $\chi^2$ found for $n=3$. The procedure was performed separately for both sets of Ramsey schemes. In Fig.\,\ref{fig:Poly_RamsS}, the result of the 50-600-50\,ms data is shown as an example, resulting in a reduced $\chi^2$ of $1.08$. In case of the 50-600-50\,ms data, the $\chi^2$ of the polynomial was $1.10$. The mass-dependent effect is more than one order of magnitude smaller than the achieved statistical uncertainty. Finally, an additional relative systematic uncertainty is taken into account for the slight differences between the production and ion-guiding parameters of \cdhzp data and $^{110}$Pd$^{+}$. Although this uncertainty is expected to lie well below the statistical uncertainty for mass doublets, the relative residual systematic uncertainty of ISOLTRAP for frequency ratios of $\delta r/r = 8 \cdot 10^{-9}$ was taken as an upper limit \cite{Kellerbauer2003}. Table \ref{tab:Qvalues} lists the obtained frequency ratios and $Q$-values as derived from Eq.\,(\ref{eq:Qvalue}) of both measurements, their weighted averages, and the $Q$-value based on AME2003 \cite{Audi2003}. The results from the two data sets agree with each other within 0.55\,$\sigma$ and the uncertainty of the weighted average of both measurements is 0.64\,keV. This is a reduction by more  
than a factor of 17 from the literature value based on the AME2003 \cite{Audi2003}.\\
Within the same measurement campaign, not only the $Q$-value, but also the absolute masses of \cdhz  and \pdhz have been determined using carbon-clusters as reference ions. Their uncertainties were reduced by a factor of 1.7 and 5, respectively. While the mass excess $ME=-90348.84\left(1.30\right)$\,keV of \cdhz is in good agreement with the literature value, the mass excess $ME=-88333.23\left(1.34\right)$\,keV of \pdhz deviates by more than 20\,keV \cite{Audi2003}. Due to the systematic mass-dependent effect, which does not cancel using reference ions of different masses, the extracted $Q$-value of $2015.6\left(1.9\right)$ has a larger uncertainty, but agrees well with the direct measurement of the $Q$-value. \\
The  phase space of the \zbb and of the \obb and thus their half-lives are affected by the new measurement of the $Q$-value. Half-lives and phase-space factors $G$ are linked via
\begin{eqnarray}
\label{eq:T1/2-2nbb}
(T^{2\nu}_{1/2})^{-1} = G^{2\nu}(Q_{\beta\beta},Z) g_A^4 \mid m_e M^{2\nu}_{GT} \mid ^2\quad \mbox{and} \\
(T^{0\nu}_{1/2})^{-1} = G^{0\nu}(Q_{\beta\beta},Z) g_A^4 \mid M^{0\nu} \mid ^2 \mid\frac{\ema}{m_e}\mid^2\,, 
\label{eq:T1/2-0nbb}
\end{eqnarray}
where $m_e$ is the mass of the electron, $g_A$ is the axial-vector coupling constant, and $\ema$ is the effective Majorana neutrino mass. $M^{2\nu}_{GT}$ and $M^{0\nu}$ are the matrix elements of the corresponding decay modes.\\ 
The new $Q$-value
\begingroup
\begin{table}[t]
\caption{Results of the evaluation of the $Q$-value measurements of the \pdhz and \cdhz mass doublet. Second column: cyclotron frequency ratios; third column: $Q$-values with uncertainties, where the residual systematic uncertainty dominates the value by more than 70\%. First and second line: individual Ramsey data sets; third line: weighted average of both data sets; fourth line: $Q$-value based on AME2003 \cite{Audi2003}.}
\begin{ruledtabular}
\begin{tabular}{lcl}
Data & Ratio $r$  & $Q$-value / keV \\ \hline
30-840-30\,ms & 1.0000197131(89) & 2018.09(90) \\ \hline
50-600-50\,ms & 1.0000197081(89) & 2017.60(90) \\ \hline \hline
Weighted Average & 1.0000197106(63) & 2017.85(64) \\ \hline \hline
AME2003 & - & 2004(11) \\ 
\end{tabular}
\end{ruledtabular}
\label{tab:Qvalues}
\end{table}
\endgroup 
 is almost 14\,keV higher and thus, increases the phase-space of the transitions and shortens the expected half-lives of both decay modes. A general formulation of the calculation of the phase-space factors was given by Doi \textit{et al.} \cite{Doi1981}. However, in contrast to previous calculations, where an approximate expression for the electron wave function at the nucleus was used, the phase-space factors reported here are calculated by a new method \cite{Kotila} using exact Dirac electron wave functions and including electron screening by the electron cloud. The wave functions are obtained by numerically solving \cite{Salvat1995} the Dirac equation with a potential for which the electron screening is taken into account through the numerically obtained Thomas-Fermi function \cite{Esposito2002}. A nuclear radius of $R=r_0 A^{1/3}$ with $r_0=1.2$\,fm was assumed for these calculations. The possible sources of uncertainties for both decay modes are the $Q$-value, the electron screening, the nuclear radius, and in case of the neutrino-accompanied mode, the closure energy. The uncertainties coming from the $Q$-value, which is dominating the total uncertainty in case of the old $Q$-value, could be reduced by more than an order of magnitude due to the more precise measurements. A comparison between the space factors based on the new $Q$-value and the $Q$-value from AME2003 and their uncertainties is shown in table\,\ref{tab:phasespace}.\\
Using the single-state-dominance hypothesis (SSDH) \cite{Civitarese1998,Domin2005,Moreno2009}, which works well for the nearby nuclei 100Mo and 116Cd, the SSDH-half-life was calculated to be $T^{2\nu}_{1/2}=1.5\left(6\right)\times10^{20}$yr, which is in good agreement with previously predicted values \cite{Civitarese1998, Domin2005, Suhonen2011}. The uncertainty of the half-life was evaluated considering the uncertainties of the $\beta$-decay half-life and the minimal and maximal value of the $\beta$-decay and electron-capture amplitudes of the 1+ intermediate state $^{110}$Ag. In general, the SSDH half-life is independent of the value of the axial coupling constant $g_A$ \cite{Domin2005}, since it is included in the $\log{ft}$-values. Nevertheless, an axial coupling constant of a free nucleon ($g_A=1.269$) was assumed, in order to calculate the matrix element for the neutrino-accompanied double-beta decay to be $M_{GT}^{2\nu} = 0.263$\,MeV$^{-1}$.\\
To specify the required sensitivity of future \obb searches, the nuclear matrix element $M^{0\nu}$ for \pdhz was calculated using two different models: the Quasiparticle Random Phase Approximation theory (QRPA) \cite{simkovic2009} and the Microscopic Interacting Boson Model (IBM-2) \cite{Iachello2011}, using the CD-Bonn and Argonne potentials for the short-range correlations as in \cite{simkovic2009}. The resulting matrix element using QRPA with the CD-Bonn potential is $M^{0\nu}=6.5\left(9\right)$ and $M^{0\nu}=5.9\left(8\right)$ with the Argonne potential. With the IBM-2 model one gets $M^{0\nu}=4.3\left(1.3\right)$ and $M^{0\nu}=4.1\left(1.2\right)$, respectively. We assumed for these calculations $g_A=1.269$ and a nuclear radius of $R=r_0 A^{1/3}$ with $r_0=1.2$\,fm. These matrix elements have relatively large values compared to most other isotopes \cite{simkovic2009,Barea2009}.\\
A reliable estimation of the half-life of the neutrinoless double-beta decay is more complicated than in case of the neutrino-accompanied mode.
\begingroup
\begin{table}[t]
\caption{Comparison of the phase-space factors for \zbb\ and \obb\ based on $Q$-values determined from the AME2003 mass values \cite{Audi2003} and the new ISOLTRAP values. The values in the first brackets are the uncertainties caused by
the uncertainty of the Q-value, while the values in the outer right brackets are the total uncertainties.}
\begin{ruledtabular}
\begin{tabular}{lcr}
$Q_{\beta \beta}$  & G$^{2\nu}$ / yr$^{-1}$ & G$^{0\nu}$ / yr$^{-1}$ \\ \hline
2004 & $1.313\left(64\right)\left(78\right)\times 10^{-19}$  & $4.716\left(23\right)\left(33\right)\times 10^{-15}$ \\ \hline
2017.8 & $1.391\left(4\right)\left(19\right)\times 10^{-19}$  &  $4.824\left(2\right)\left(12\right)\times 10^{-15}$ \\ 
\end{tabular}
\end{ruledtabular}
\label{tab:phasespace}
\end{table}
\endgroup 
Here, the uncertainty is mainly due to the unknown neutrino mass. Using Eq.\,(\ref{eq:T1/2-0nbb}) and the weighted average $M^{0\nu}=5.5$ of the four different matrix elements, obtained with the QRPA- and IBM-2-model and the CD-Bonn and Argonne potentials, the half-life can be estimated by 
\be
T_{1/2}^{0\nu} = \frac{6.8 \times 10^{23}\, \mbox{eV}^2\,\mbox{yr}}{\mid\ema\mid^2}\quad.
\label{eq:T1/2-0nbb-appr}
\ee 
Assuming the effective neutrino mass in the range from 1\,eV down to $10^{-3}$eV leads to a \obb half-life between $6.8 \times 10^{23}$\,yr and $6.8 \times 10^{29}$\,yr, respectively. Currently, a possible evidence for \obb in $^{76}$Ge is discussed resulting in an effective Majorana neutrino mass of $\ema=0.32\left(3\right)\,$eV \cite{klapdor2006}. To explore this mass range, a \obb experiment on \pdhz must be sensitive to a half-life range from $T^{0\nu}_{1/2}=5 \times 10^{24}$\,yr to $T^{0\nu}_{1/2}=1 \times 10^{25}$\,yr.\\
In conclusion, the $Q$-value of the double-beta transition from \pdhz to \cdhz was measured for the first time by high-precision Penning-trap mass spectrometry and resulted in a value of $Q=2017.85(64)$\,keV. Thus, the uncertainty was reduced by a factor of 17 with respect to the AME2003 value, revealing a 14\,keV higher $Q$-value compared to the literature value. In addition, masses of \pdhz and \cdhz were determined, reducing their uncertainties significantly. Based on the more accurate $Q$-value, the phase-space factors of the \zbb and the \obb were recalculated by a new, more precise method. The SSDH half-life calculation leads to an expected half-life of the neutrino-accompanied double-beta decay of $T^{2\nu}_{1/2}=1.5\left(6\right)\times 10^{20}$\,yr. In case of the neutrinoless double-beta decay, the matrix elements have been calculated with the help of the QRPA and IBM-2 models, resulting in large values compared to most other possible double-beta decay nuclides. Thus, in combination with its high natural abundance, \pdhz becomes a very promising candidate for double-beta decay studies and for the search for the neutrino mass.  
\begin{acknowledgments}
The authors thank Dr.\,Francesco Iachello for his contribution to the calculation of the phase-space factors and matrix elements. This work was supported by the German Federal Ministry for Education and Research (BMBF) (06DD9054, 06GF186I), by the Max-Planck Society, by the French IN2P3, by the DIUC project no. 211.011.055-1.0, Universidad de Concepci\'on, Chile, and by the ISOLDE Collaboration.  
\end{acknowledgments}

\end{document}